\documentclass[pr,nofootinbib,twocolumn,showpacs,aps]{revtex4-1}
\usepackage{bm}
\usepackage{graphicx}
\usepackage{amssymb}
\usepackage{amsmath}
\usepackage{eufrak}
\usepackage{color}
\usepackage[utf8]{inputenc}
\usepackage[unicode=true,colorlinks=true,urlcolor=blue,citecolor=blue]{hyperref}

\usepackage{pifont}
\usepackage{ulem}

\begin{document}

\title{Manifestation of interface anisotropy in CdTe quantum wells}

\author{L.\,V.\, Kotova$^{1,2}$}
\author{A. V. Platonov$^1$} 
\author{R. Andr\'e$^3$} 
\author{ H. Mariette$^3$}
\author{V. P. Kochereshko$^1$}

\affiliation{$^1$Ioffe Institute, 194021 St.~Petersburg, Russia}
\affiliation{$^2$ITMO University, 197101 St.~Petersburg, Russia} 
\affiliation{$^3$ Institut Neel, CNRS, F-38000 Grenoble, France} 


\begin{abstract}
Photoluminescence and polarized reflection spectra of quantum well structures with symmetric  Cd$_{0.9}$Zn$_{0.1}$Te/CdTe/Cd$_{0.9}$Zn$_{0.1}$Te and asymmetric Cd$_{0.9}$Zn$_{0.1}$Te/CdTe/Cd$_{0.4}$Mg$_{0.6}$Te barriers were studied. The Stokes parameters of the reflected light from these structures were measured. In the structures with symmetric barriers, exciton resonances were found in the reflection spectra and were not present in the photoluminescence spectra. In structures with asymmetric barriers, in the region of exciton resonances, the phenomenon of light birefringence was detected, caused by a lower symmetry of the interfaces compared to the symmetry of bulk crystals. A discussion of both phenomena was given.

\end{abstract}

\maketitle{}

\section{Introduction}
    A$_{2}$B$_{6}$ semiconductor compounds, particularly CdTe and ZnTe, are often used for basic research. Unfortunately, the use of their heterostructures for device applications is not very wide due to rapid degradation.
However, cadmium telluride has become a verified thin film solar cell material due to its unique optical properties, namely a direct bandgap with a high absorption coefficient. The exploration of CdS/CdTe heterojunction solar cells started more than thirty years ago and the current efficiency of the CdTe solar cell has reached 22.1~\%, the leading CdTe thin film-based photo voltaic manufacturing company (see Ref.~\cite{1,book} and all the ref. therein).

One of the reasons preventing the vide practical use of heterostructures based on CdTe and ZnTe compounds is the noticeable mismatch of their crystal lattices (6.4~\%). As a result, mechanical stresses arise at the interfaces, which can lead to a phase transition in CdTe/ZnTe strained layer superlattices~\cite{2}. Moreover, due to lattice mismatch, the magnitude of the band offset is strongly dependent on the strain~\cite{3}. As a result, the value of the band offset is known rather approximately. For example, the scatter of published data on the band discontinuity in the valence band of CdTe/ZnTe heterostructures is up to 10~\% of the total band gap~\cite{2,3}. 

Those who manufacture heterolasers empirically know  that a "good" laser will be obtained if it is fabricated, for example, in the direction [110] and "bad” - is fabricated in the orthogonal direction $[1\bm{\bar}{1}0]$. This is partly due to the fact that these directions correspond to the directions of dislocation creation. Knowing that the interfaces are the place of maximum mismatch of the crystal lattices of the materials in contact and, consequently, the place of the dislocation origin, the study of interface properties is therefore of crucial importance for practical task.

In this paper a detailed experimental study of photoluminescence and polarized reflection spectra from structures with quantum wells with symmetric and asymmetric barriers was carried out. A birefringence phenomenon caused by reduced interface symmetry in the heterostructure was found in the structure with asymmetric barriers. Previously, the manifestation of reduced interface symmetry was observed in the photoluminescence spectra of type-II heterostructures, where the exciton is "bound" directly to the interface~\cite{4, 5, 6, 7} that is in local properties of the structures. In contrast to these publications, in this study, the reduced interface symmetry was manifested in the dielectric response of type-I structures near exciton resonances.

The paper was organised as follows. In Sec.~\ref{exp} we described our experiments. Sect.~\ref{theory} gave a microscopic theory of excitons taking into account strain in symmetric and asymmetric quantum wells. In Sec.~\ref{Disc} experimental results were compared with calculations which allow to identify lines in spectra. Also, the cause of polarization conversion in the asymmetric structure was discussed and experimental results were compared with phenomenological theory. Concluding remarks were given in Sec.~\ref{Concl}.
 \begin{figure*}
\centering
 \includegraphics[width=2.05\columnwidth]{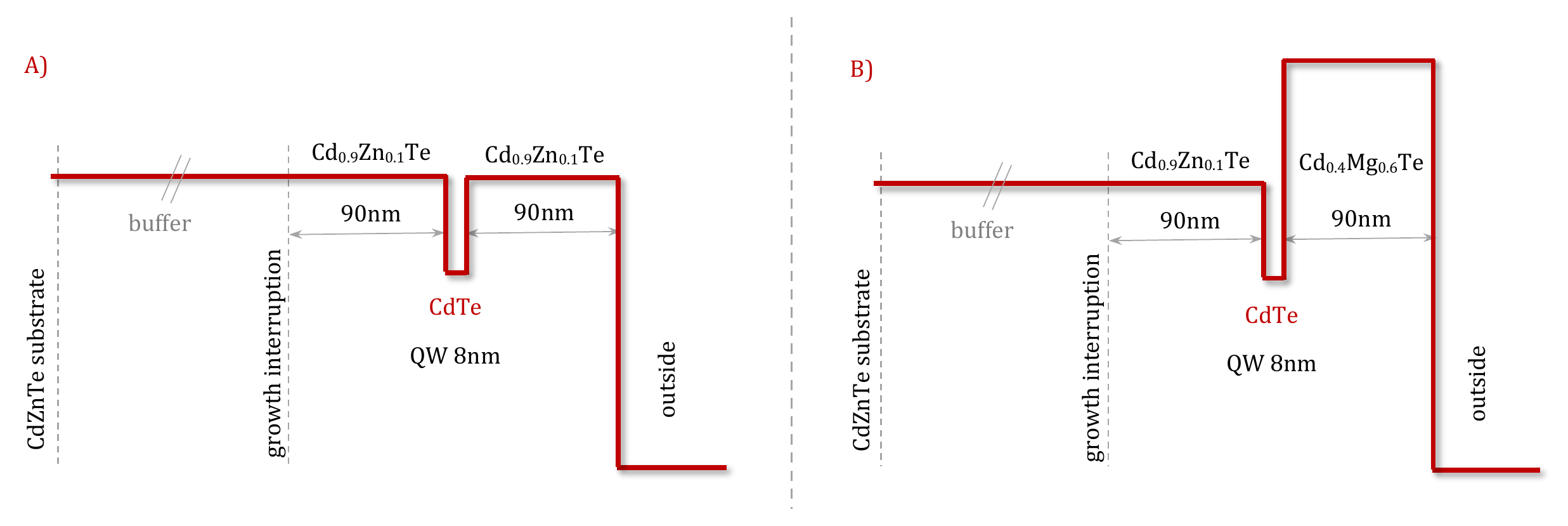}
  \caption{Scheme of the investigated structures. (A) Structure with a single CdTe quantum well of 8~nm width and Cd$_{0.9}$Zn$_{0.1}$Te symmetric barriers of 90 nm width. (B) Structure with a single CdTe quantum well of 8~nm width and asymmetric Cd$_{0.9}$Zn$_{0.1}$Te and Cd$_{0.4}$Mg$_{0.6}$Te barriers of 90~nm width each. }
  \label{fgr:Sample}
 \end{figure*}
 
\section{Experiment}
\label{exp}
CdTe/Cd$_{0.9}$Zn$_{0.1}$Te based structures with single 8 nm wide quantum well grown by molecular beam epitaxy in the [001] direction were investigated. Reflection high-energy electron diffraction was used to optimize the two-dimensional layer by layer growth of CdTe and the Cd$_{1-x}$Zn$_{x}$Te alloy and to measure the layer thickness to one monolayer accuracy~\cite{8}. A set of such structures with symmetric and asymmetric barriers was fabricated. In the first case the quantum well was surrounded by symmetric barriers with 10~\% Zn composition on both sides, and in the second case one of the barriers was the same as in the symmetric structure, and the other barrier was based on Cd$_{0.4}$Mg$_{0.6}$Te (Fig.~\ref{fgr:Sample}~a,b).
The height of these barriers differed by more than a factor of two. The structure parameters are given in Figure~\ref{fgr:Sample}. Both types of structures, with symmetric and asymmetric barriers, were grown on Cd$_{1-x}$Zn$_{x}$Te substrates with composition x=4~\% and x=20~\%. The use of structures with a large Zn composition in the substrate made it possible to record not only photoluminescence (PL) and reflection spectra, but also transmission spectra. 

The spectra were taken using a 0.5~M monochromator and recorded with an CCD detector. A halogen lamp was used as a light source to registered transmission and reflection spectra, and a laser with wavelength of 533~nm was used for the PL spectra. 

\begin{figure}[b]
\centering
 \includegraphics[width=0.8\columnwidth]{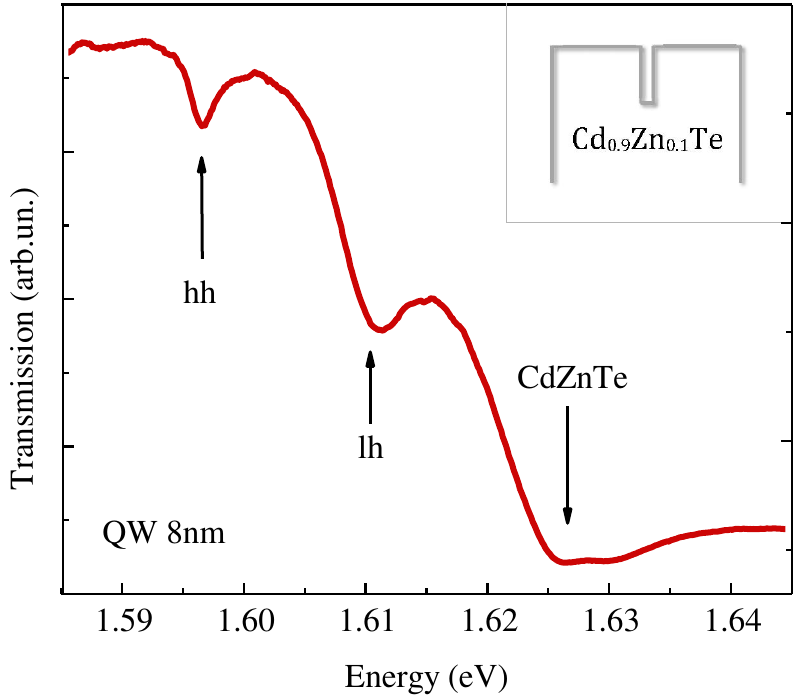}
  \caption{ Transmission spectrum of the structure with a single CdTe quantum well of 80~A width with symmetric Cd$_{0.9}$Zn$_{0.1}$Te barriers, taken at 77~K. The feature at energy 1.595~eV is due to heavy hole exciton absorption. The feature at energy 1.611~eV is related to light hole exciton absorption. The minimum at energy 1.62~eV is associated with absorption in Cd$_{0.9}$Zn$_{0.1}$Te barriers. }
  \label{fgr:Trans}
 \end{figure}

Figure~\ref{fgr:Trans} shows the transmission spectrum of the sample with a symmetric quantum well, with 20~\% Zn in the substrate, taken at 77~K. The transmission spectrum of the asymmetric sample qualitatively coincided well with this spectrum. Structures grown on substrates with a Zn composition of 4~\% fully absorbed light at exciton resonance energies in the quantum well. However, the reflection spectra carried the same information as the transmission spectra and may well characterize these structures. 
In the transmission spectrum, at energy 1.595~eV a peculiarity associated with absorption to the heavy exciton states in the quantum well is visible. At energies of 1.626~eV or less, absorption is observed in the Cd$_{0.9}$Zn$_{0.1}$Te barrier layer. At an energy of 1.610~eV a feature related to the light exciton in the quantum well can be seen. 

\begin{figure*}
  \includegraphics[width=0.86\columnwidth]{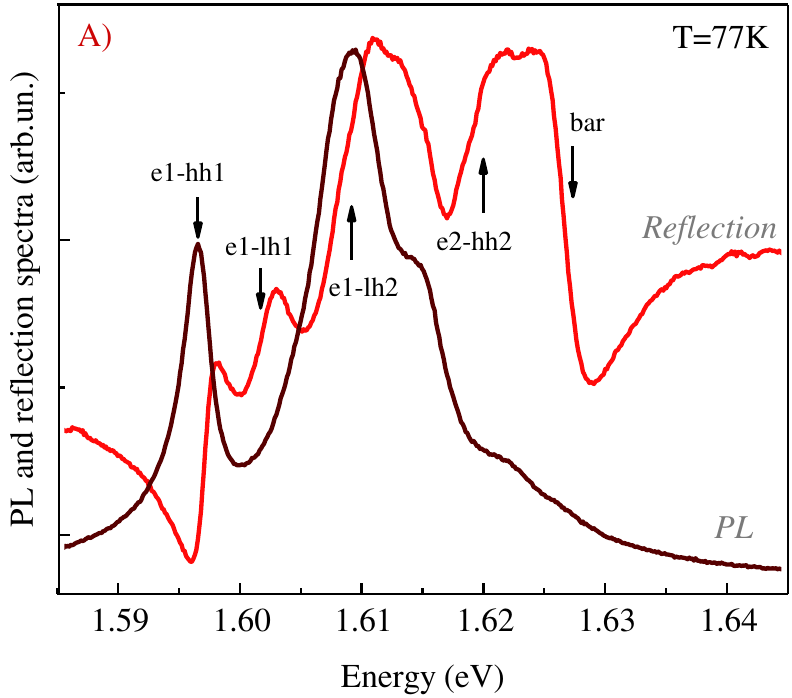}
  \includegraphics[width=0.8\columnwidth]{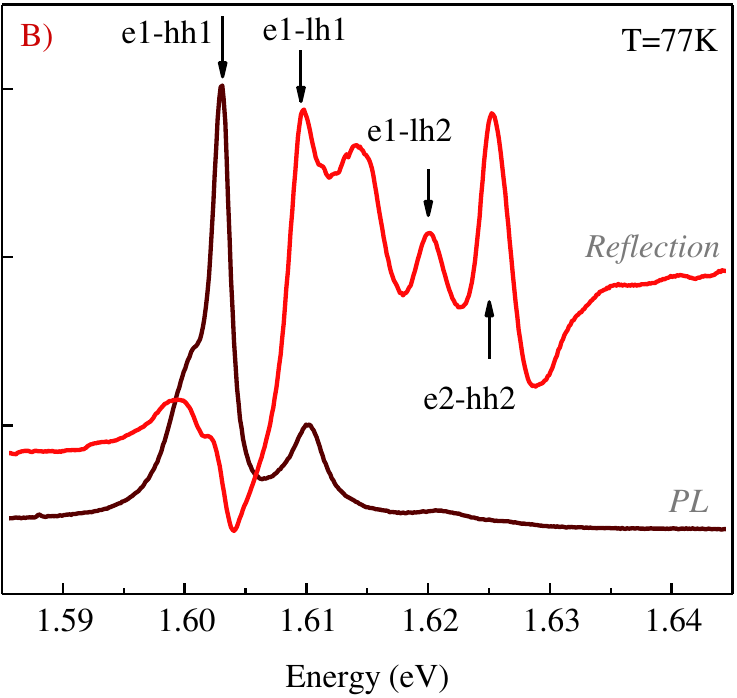}
  \caption{ Reflection and photoluminescence spectra of structures with quantum wells. (A) - with symmetric and (B) - with asymmetric barriers taken at 77~K. The arrows indicate the optical transitions to exciton states with heavy and light holes. The identification of these transitions was performed on the basis of calculations and described in detail in the discussion Sec.~\ref{Disc}.  }
  \label{fgr:RefPL}
 \end{figure*}

	Figure~\ref{fgr:RefPL} shows the reflection and photoluminescence (PL) spectra of the structures with symmetric and asymmetric barriers. The reflection spectra show many more peculiarities than the absorption spectra. The main reason of it is, due to the small thickness, the exciton absorption is also small, and only states with a sufficiently large oscillator strength are visible in the absorption spectrum. The Stokes shift between the spectral features in the reflection and PL spectra is small, indicating a high quality of the structures. What is striking is that the reflection and PL spectra show too many lines for a structure with a single quantum well. 
	
Comparing the reflection and the PL spectra of symmetric structures, we can see that some resonance features are present in the reflection spectra and absent in the PL. Usually, the opposite is the case when the lines are present in the PL spectra and absent or weak in the reflection spectra, which is associated with relaxation of carriers downward in energy. In the structure with symmetric barriers (Fig.~\ref{fgr:RefPL} a), there is a bright feature in the reflection spectrum at an energy of 1.618 eV that is almost absent in the PL. This can only be explained by the fact that at non-resonant photoexcitation the carriers cannot populate this state, but the oscillator strength of the direct optical transition to this state is large. 

The most intense line in the PL spectrum is the line at energy 1.609~eV, the intensity of the other features is remarkably lower. This indicates that relaxation into lower energy states is difficult from this state, and all photoexcited carriers emit through these state. 

For the initial identification of the lines in the reflection spectrum, we used the result of ref.~\cite{9} based on the shape of the exciton reflection line. In this paper, it was shown that the shape of the exciton reflection line is determined by the light interference reflected from the surface and the quantum well. Depending on the phase shift ($\varphi=kd, k$ is wavevector) for the light passes the distance from the surface to the quantum well and back, the shape of the spectrum can change significantly. The reflection spectrum (Fig.~\ref{fgr:RefPL} a) shows that the exciton reflection contour from the surface of the structure, at energy 1.626~eV has a "differential" shape with a maximum at energy 1.624~eV and a minimum at energy 1.628~eV. The features of the reflection spectra at energies 1.597~eV, 1.602~eV, 1.609~eV, and 1.620~eV have an "opposite" shape with a minimum at low energies and a maximum at higher ones. Following ref.~\cite{9}, these features should be related to the reflection from the quantum well.

A similar procedure was performed to identify the spectra of the asymmetric structure (Fig.~\ref{fgr:RefPL} b). This allowed us to identify the spectral features at energies 1.604~eV, 1.610~eV, 1.620~eV, and 1.625~eV as belonging to the quantum well. 

Polarized reflection spectra at normal and oblique light incidence from these structures were also measured. At oblique incidence of light, in both structures, the phenomenon of polarization conversion from linear to circular was observed. This phenomenon is related to the gyrotropy of quantum wells~\cite{Opt_act_PRB} and was studied in detail in our previous works~\cite{10, 11}. No polarization conversion was observed in the structure with symmetric barriers at normal light incidence~\cite{11}. The degree of circular polarization was only a few percent at oblique light incidence~\cite{11}. 

\begin{figure*}
\centering
 \includegraphics[width=0.85\columnwidth]{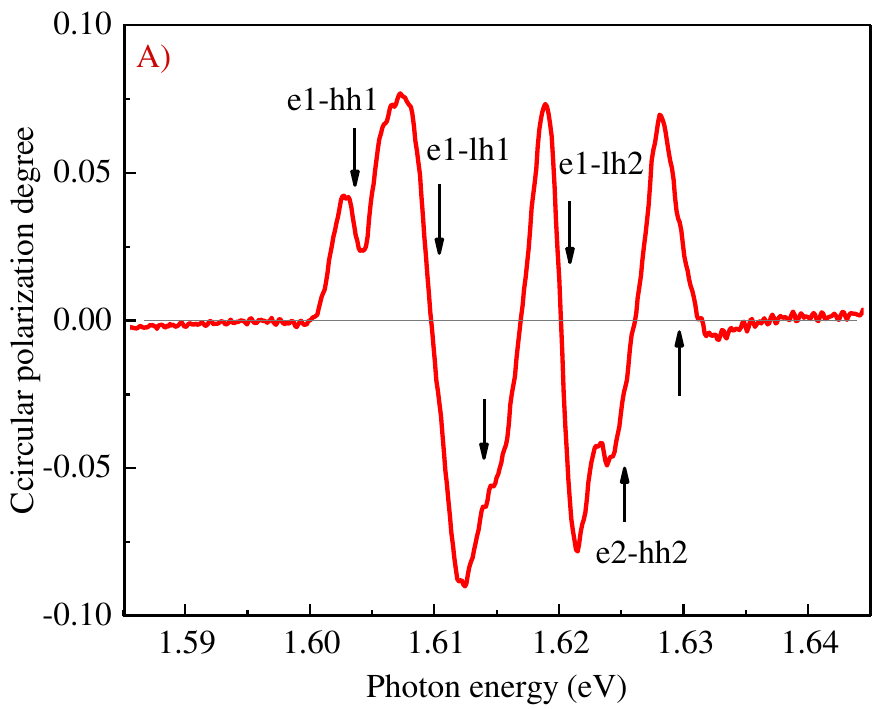}
  \includegraphics[width=0.85\columnwidth]{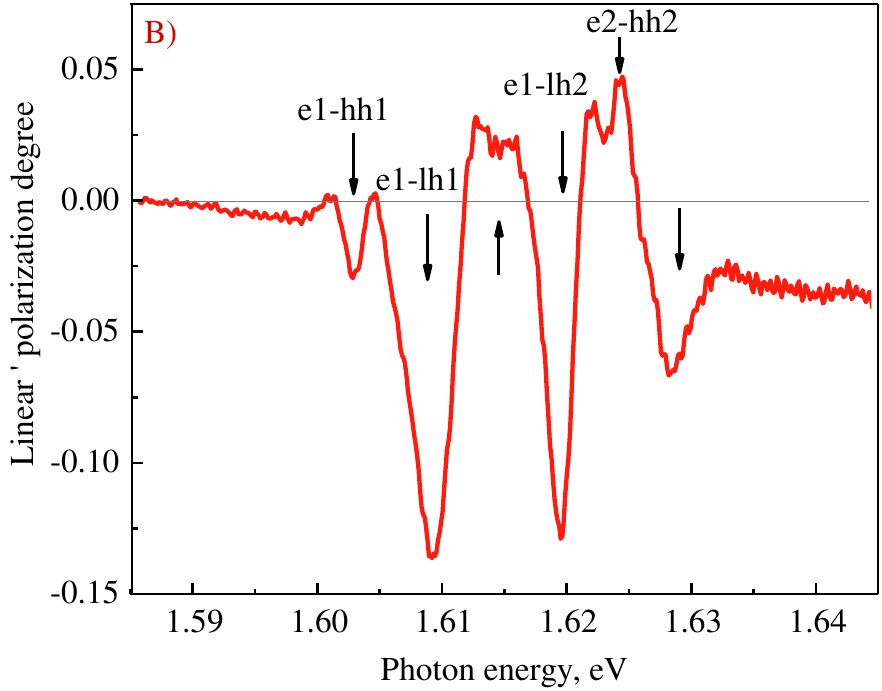}
  \caption{ Polarization degree of linearly polarized in the direction [100] light reflected from the structure with asymmetric barriers at normal incidence. A) degree of circular polarization, B) degree of linear polarization in the axes rotated by 45~degrees relative to the polarization of incident light. The arrows mark the features observed in the reflection spectrum (Fig.~\ref{fgr:RefPL} b). }
  \label{fgr:conver}
 \end{figure*}
 
At the same time, in the structure with asymmetric barriers, a large value of conversion, reaching tens of percent, was found even at the normal light incidence (Figure~\ref{fgr:conver}). The maximum value of circular polarization in the reflection spectrum was observed when the incident light was polarized along the [100] or [010] axis. If the incident light was polarized along the [110] or $[1\bm{\bar}{1}0]$ axes, the circular polarization of the reflected light did not exceed 2~\%.

Fig.~\ref{fgr:conver} a shows the spectral dependence of the degree of circular polarization of light $P_{\rm circ}$ and Fig.~\ref{fgr:conver} b - the linear polarization ${P}_{\rm lin'}$ in the axes rotated by 45~degrees relative to the polarization of incident light. 

\begin{equation}
\label{S}
P_{ \rm circ}={R^{\sigma{}^+}-R^{\sigma{}^-} \over R^{\sigma{}^+}+R^{\sigma{}^-}}, 
	\quad 
	{P}_{ \rm lin'}={{R}^{x'}-{R}^{y'} \over {R}^{x'}+{R}^{y'}} . 
	\nonumber
\end{equation}

The spectral features of these dependences coincide with the spectral features of the reflection spectra, as indicated by arrows in Figure~\ref{fgr:conver}. Note that the minima in the dependence ${P}_{ \rm lin'}(\omega)$ coincide with the zeros in the dependence $P_{ \rm circ}(\omega)$.

Conversion of the linear light polarization to circular polarization in this case was directly related to the birefringence phenomenon with the optical axis along [110] or $[1\bm{\bar}{1}0]$]. However, despite the fact that the investigated structures were really strained, this could not lead to the appearance of optical axis in the plane of the structure. Indeed, birefringence emerged only in the asymmetric structure and was absent in the symmetric structure having the same mechanical stresses. It was natural to assume that the birefringence was due to the asymmetry of the interfaces. In~\cite{7} it had been shown with ZnSe/BeTe type-II quantum wells that in QWs based on cubic semiconductor with $D_{2d}$ symmetry, each interface had $C_{2v}$ symmetry. Such reduced symmetry could lead to interface birefringence of light. However, one wondered how an interface layer with a thickness on the order of a lattice constant could lead to such a strong effect and what role excitons played here. 

To reliably identify the observed optical transitions, we performed theoretical calculations of the energies of the excitonic states. 
\section{Theory}
\label{theory}
Let us estimate the positions of the lines of light and heavy hole excitons in this structure. The lattice constants in the materials CdZnTe and CdTe differ by about 0.7~\%. As a result, in addition to the potential associated with the so-called chemical band offset between the contacting materials, the potential associated with the strain band offset acts on the carriers. The value of the chemical band offset in the valence band of CdTe-CdZnTe heterostructures is known very approximately and is considered to lie in the range from -10~\% to +10~\% of the total band gap~\cite{3, 7}.

The value of the total chemical band offset is equal to the difference between the band gaps of the bulk materials Cd$_{0.9}$Zn$_{0.1}$Te and CdTe: $\Delta={E_g}^{ZnTe}-{E_g}^{CdTe}=0.063$~eV~\cite{13, 14, 15}. This value is divided between the conduction band ($\Delta_c$) and the valence band ($\Delta_\nu$), $\Delta=\Delta_c+\Delta_\nu$  . In this work, we assume $\Delta_c=0.97\Delta$ and $\Delta_\nu=0.03\Delta$~\cite{16}. Since the buffer layer and barrier layers are much thicker than the quantum well layer, we can assume that these layers are not stressed and all mechanical stresses are concentrated in the well. The strain discontinuity of the bands can be calculated by formulas~\eqref{f1}~\cite{17, 18}: 

\begin{eqnarray}
\label{f1}
\Delta E_c=2a_c (S_{11}+2S_{12})\sigma,
\nonumber \\
\Delta E_{hh}=2a_\nu (S_{11}+2S_{12})\sigma+b(S_{11}-S_{12})\sigma,
\nonumber \\
\Delta E_{lh}=2a_\nu (S_{11}+2S_{12})\sigma-b(S_{11}-S_{12})\sigma.
\end{eqnarray}
Here $\Delta E_c$  is deformational band offset in the conductivity band, $\Delta E_{hh}$  is deformational band offset in the heavy hole band,  $\Delta E_{lh}$ is deformational band offset in the light hole band, $a_c$  and  $a_\nu$ are hydrostatic deformational potentials in the conduction and valence bands,  $b$ is uniaxial deformation potential, $S_{ij}\times10^{-11}  m^2/N$ are elastic constants,  $\sigma = {\varepsilon / (S_{11}+S_{12})}$ is mechanical in-plane tension, $\varepsilon=({a_j}^L - {a_i}^L) / {a_i}^L$  is in-plane deformation. 

For the deformation potentials (see Table~\ref{tbl1}) the relations $a~=~a_c~-~a_\nu,  {a_c / a_\nu} = -2 $  are valid~\cite{17}. 
{\begin{table} [h]
\small
  \caption{\ Deformation potentials and elastic constants in CdTe.}
  \label{tbl1}
 \begin{tabular*}{\linewidth}{@{\extracolsep{\fill}}cccccc}
    \hline
    $S_{11}$ & $S_{12}$ &  $S_{44}$ &  $a$  [eV] &  $b$  [eV] & $\sigma $ \\
    \hline
    3.581  & -1.394  & 5.5  &-3.85 &-1.20 & 0.0358 \\
      \hline
 \end{tabular*}
\end{table}}

Using formula~\eqref{f1}, we obtain that the strain band offset in the conduction band = 15~meV, the strain band offset in the heavy hole subband  = 2.1~meV, the strain band offset in the light hole subband  = -16.5~meV.

The deformation band gap (Tables~\ref{tbl2} and \ref{tbl3}) is summed with the chemical band gap so that for light holes we get a type-II heterostructure due to the deformation, and for heavy holes we get a type-I structure.

{\begin{table} [h]
\small
  \caption{\ Bandgaps in CdTe, CdZnTe and CdMgTe~\cite{13} at 77K.}
  \label{tbl2}
 \begin{tabular*}{\linewidth}{@{\extracolsep{\fill}}cccc}
    \hline
    CdTe  & CdTe deform. &  Cd$_{0.9}$Zn$_{0.1}$Te & Cd$_{0.4}$Mg$_{0.6}$Te \\
    \hline
    1.576 eV  & 1.589 eV & 1.636 eV  &2.70 eV  \\
      \hline
 \end{tabular*}
\end{table}}

{\begin{table} 
\small
  \caption{\ Total (chemical+strain) band discontinuities in CdTe/CdZnTe and CdTe/CdMgTe heterostructures~\cite{13}. }
  \label{tbl3}
 \begin{tabular*}{\linewidth}{@{\extracolsep{\fill}}cccccc}
    \hline
     & CdTe -  &  Cd$_{0.9}$Zn$_{0.1}$Te &   &   CdTe -  & Cd$_{0.4}$Mg$_{0.6}$Te\\
    \hline
    CBO  & VBOhh  & VBOlh  &CBO &VBOhh & VBOlh \\
     \hline
    45 meV & 2 meV	 &-12.5 meV	 &1.5 eV &	0.37 meV &	0.37 meV \\
      \hline
 \end{tabular*}
\end{table}}

Since the value of the total band gap in the valence band of the studied structure is small, the Coulomb interaction with the electron becomes the main contribution to the potential energy of holes~\cite{19}.

We assume that the electron and hole are quantized in the potential wells for the electron and hole formed as a result of a band offset between the CdTe and CdZnTe layers and bound to each other by Coulomb interaction. The Schrödinger equation for the exciton in this case has the form: 

\begin{widetext}
\begin{multline}
\label{f2}
\left[ \,-{{\hbar}^2 \over 2m_e } {{\partial}^2 \over {	\partial {z_e}^2} } + V (z_e)-{{\hbar}^2 \over 2m_h } {{\partial}^2 \over {	\partial {z_h}^2} } + V (z_h) - {{{\hbar}^2 \over 2\mu }
\left({1 \over \rho } {\partial\over {\partial\rho}}\rho {\partial\over {\partial\rho}} + {1 \over \rho^2}{\partial^2\over {\partial{\phi}^2}}
\right)} - {{e^2} \over {\varepsilon_0 \sqrt{\rho^2+{\vert z_e -z_h\vert}^2}}}\right] \,\Psi(\overrightarrow{r_e},\overrightarrow{r_h}) \\
= \left(E- {{{\hbar}^2 Q_\bot^2}\over 2M }\right)\Psi(\overrightarrow{r_e},\overrightarrow{r_h}), 
\end{multline}
here:
\begin{equation*}
\phi=\arctan \left({{x_e-x_h}\over {y_e-y_h}}\right), \rho = \sqrt{{\vert{x_e-x_h}\vert}^2+{\vert{y_e-y_h}\vert}^2},
\nonumber
\end{equation*}
$V (z_e)$  is rectangular potentials for electrons and $ V (z_h)$  is for holes,  $E$ is total exciton energy, $M$   is the exciton mass, $Q_\bot$  is in-plane center of mass wavevector. 
\end{widetext}
For an approximate solution of equation~\eqref{f2}, we will use the idea of the paper~\cite{19}. However, unlike this paper, in our case the Bohr radius of the exciton is much smaller than the width of the quantum well. 

As in~\cite{19}, we assume that the kinetic energy of the electron in the quantum well is much greater than the exciton binding energy ({\it adiabatic approximation}):
\begin{equation*}
{{\hbar}^2 \over 2m_e L^2 } \gg {e^2 \over \varepsilon_0\tilde{a}_B},
\nonumber
\end{equation*}
$L$ is quantum well  $V (z_e)$ width,  $\tilde{a}_B$ is exciton Bohr radius in the quantum well plane.

We first need to solve the equation of the electron in the quantum well  $V (z_e)$ in $z$  direction, and then solve the equation of motion of the hole in the average potential created by the electron and the potential of the quantum well  $ V (z_h)$. For the electron: 

\begin{equation}
\label{f3}
\left[ \,-{{\hbar}^2 \over 2m_e } {{\partial}^2 \over {	\partial z_e^2} } + V (z_e)-E_e\right]\,\phi_e(z_e)=0.  
\end{equation}

We find the quantization energy along the z-axis and the wave functions of the electron along the z-axis. Neglecting the tails of the wave function in the barriers we obtain: 
\begin{equation}
\label{f4}
\phi_e(z_e)=\sqrt{2 \over L} \cos \left({\pi \over L}z_e\right).  
\end{equation}

The second step is to solve the equation for the two-dimensional exciton, which is: 
\begin{equation}
\label{f5}
\left[ \, - {{{\hbar}^2 \over 2\mu }\left({1 \over \rho } {\partial\over {\partial\rho}}\rho {\partial\over {\partial\rho}} + {1 \over \rho^2}{\partial^2\over {\partial{\phi}^2}}\right)} - {{e^2} \over {\varepsilon_0 \rho}} - E^{exc}_{2D}\right]\, f_n (\rho,\phi)=0.
\end{equation}
Solution for the ground state of the exciton: 
\begin{equation}
\label{f6}
 f_1 (\rho,\phi)=\sqrt{2 \over \pi} {2 \over a_B} e^ {-{2\rho / a_B }} = \sqrt{2 \over \pi} {1 \over \tilde{a}_B } e^ {-{\rho / \tilde{a}_B }} ,
\end{equation}
where   $\tilde{a}_B= {a_B / 2}$ is two-dimensional exciton Bohr radius, $a_B$  is bulk exciton Bohr radius. 

\begin{equation}
\label{f7}
a_B= {\varepsilon_0 {\hbar}^2 \over \mu e^2 }=\sqrt{{\hbar}^2 \over 2\mu E^{exc}_{3D}}.
\end{equation}
Exciton Rydberg for 2D exciton is \begin{equation}
E^{exc}_{2D} = - {\mu e^4 \over 2 {\hbar}^2 \varepsilon_0^2}{1 \over (n+ 1/2)^2}.
\nonumber
\end{equation} 
Thus, the effective potential acting on the hole is: 

\begin{widetext}
\begin{equation}
\label{f8}
    V_{eff}(z_h)= V(z_h) - R y^{2D} - {e^2 \over \varepsilon_0} {2 \over L}{4 \over \tilde{a}^2_B }\int\limits_{L/2}^{-L/2} \,dz_e { \left | \cos{\left ({\pi \over L} z_e\right)} \right |}^2 \int\limits_0^\infty \rho\,d\rho e^{-{2\rho / \tilde{a}_B }} \left[ \, - {1 \over \rho} + {1 \over \sqrt{(\rho^2 + (z_e-z_h)^2}}  \right]\,.
\end{equation}

Let us decompose the root, assuming $(z_e-z_h) \gg \rho$  . Denote $y\equiv z_e-z_h, k = \pi / L$   get it: 
\begin{equation}
\label{f9}
 V_{eff}(z_h)= V(z_h) + {e^2 \over \varepsilon_0} {2 \over \tilde{a}_B}+ {e^2 \over \varepsilon_0} {2 \over L}{4 \over \tilde{a}^2_B }\int\limits_{L}^{-L} \,dy  { \cos^2{\left \lbrace{\pi \over L} (y+z_h)\right \rbrace} } \sum_{n=1}^\infty \int\limits_0^\infty \,d\rho e^{-{2\rho / \tilde{a}_B }} {\rho \over {\vert y \vert}} \left( \, 1+ (-1)^n {(2n-1)! \over 2n!} \left({\rho \over  y }\right)^{2n}  \right)\,.
\end{equation}

Integrating by  $\rho$ and by $y$  we get:

    \begin{equation}
\label{f10}
 V_{eff}(z_h)= V(z_h) + {e^2 \over \varepsilon_0} {2 \over \tilde{a}_B}+ {e^2 \over \varepsilon_0} {4 \over \tilde{a}^2_B } {L \over \pi}\left( \sum_{n=1} (-1)^n {(2n-1)!(2n+1) \over 2n\pi^{2n}}  \left( {\tilde{a}_B \over 2} \right)^{2n+1} \left( {\pi \over L} \right)^{2n+1}  \left( 1+ {cos^2\left( {\pi \over L} z \right) \over (2)^{2n}} \right)\right)\,.
\end{equation}

Limiting ourselves to the first two terms: 

    \begin{equation}
\label{f11}
V_{eff}(z_h)\approx V(z_h) + {e^2 \over \varepsilon_0} {2 \over \tilde{a}_B} - {e^2 \over \varepsilon_0 \tilde{a}_B} {3 \over 2 } \left({ \tilde{a}_B \over L}\right)^2\left( \, 1+ {1 \over 4}cos^2\left( {\pi \over L} z_h \right) \right)\,.
\end{equation}

\end{widetext}

The lower quantization level of a hole in such a potential, counted from the bottom of the hole band: 
\begin{equation}
\hbar\omega={\tilde{a}_B \over L}\sqrt{{3\over 4}{e^2 \over \varepsilon \tilde{a}_B}{\pi^2 {\hbar}^2 \over {m_h}L^2}} \approx10.5~meV .
\nonumber
\end{equation}

\begin{figure*}
\centering
 \includegraphics[width=0.85\columnwidth]{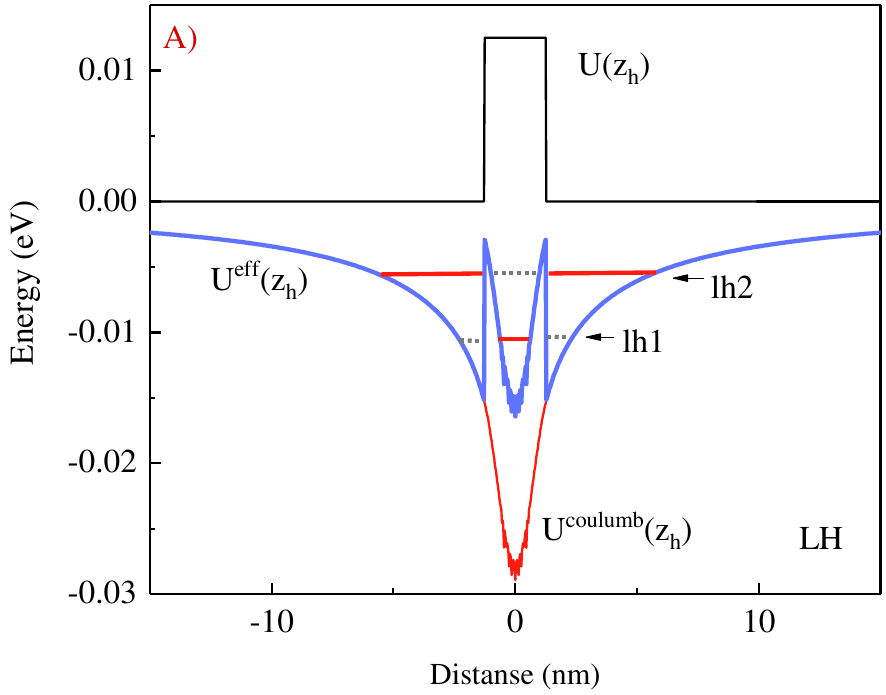}
  \includegraphics[width=0.85\columnwidth]{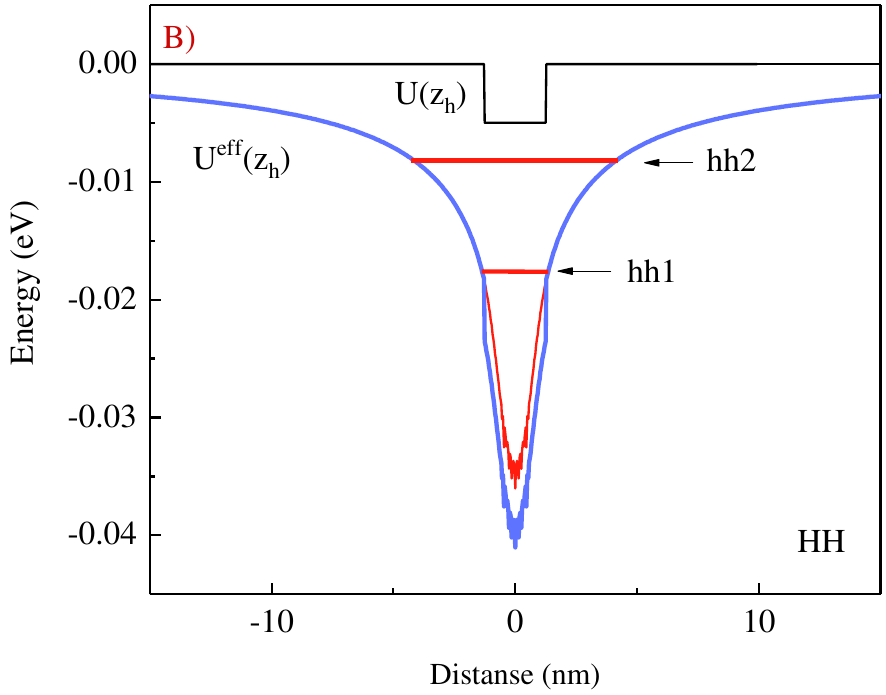}
  \includegraphics[width=0.85\columnwidth]{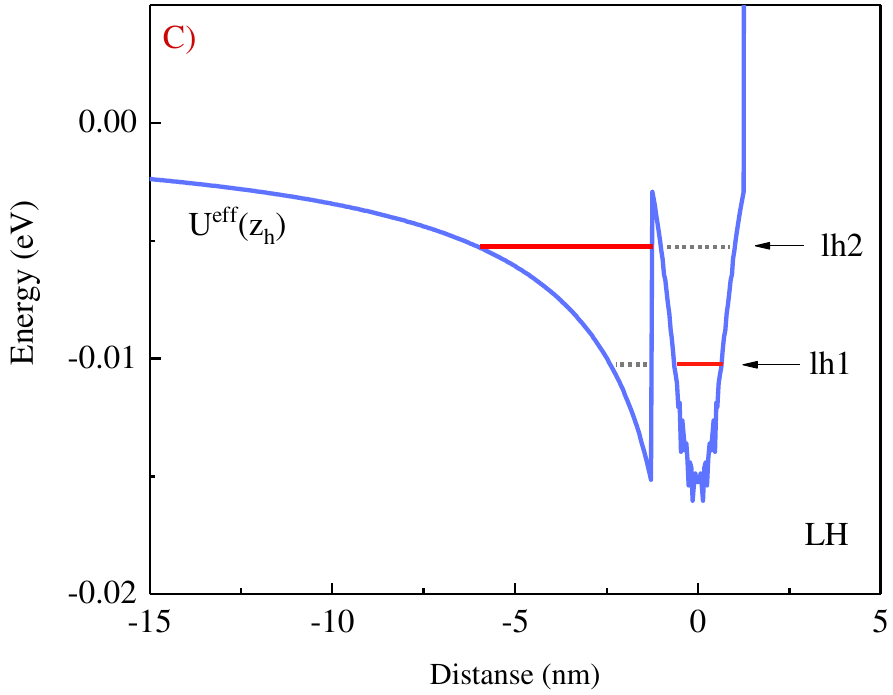}
  \includegraphics[width=0.85\columnwidth]{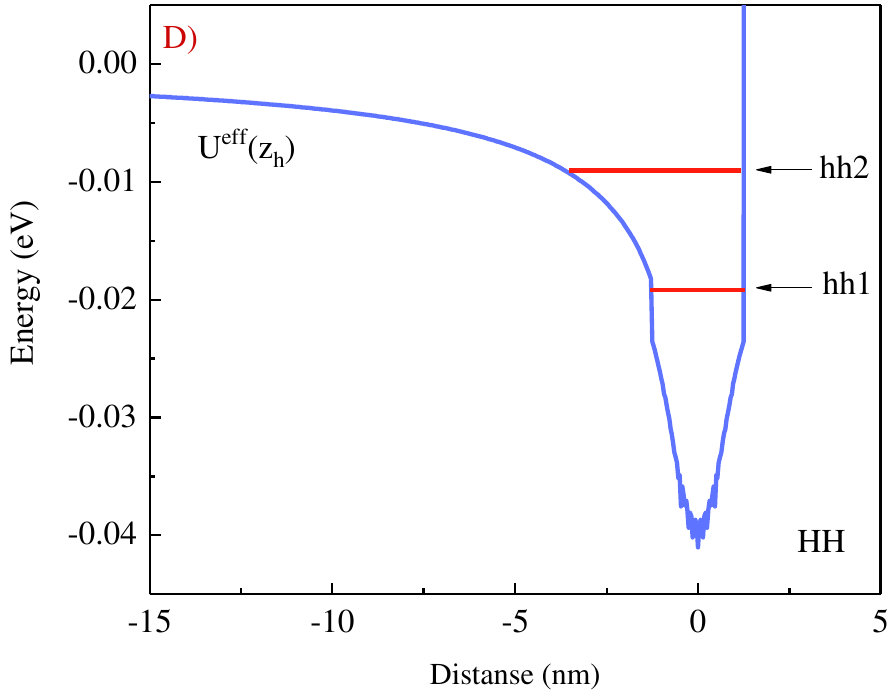}
  \caption{ Calculation of the effective potential for hole motion created by the electron in the quantum well, dot line means hole can tunnel their.  $U (z_h)$ are nominal potentials without corrections for heavy and light holes. $U^{(coulumb)} (z_h)$  is averaged Coulomb potential.  $U^{eff} (z_h)=U (z_h) + U^{(coulumb)} (z_h)$ is effective potential acting on the hole. A) and C) are for light holes in symmetric and asymmetric wells respectively, B) and D) are for heavy holes in symmetric and asymmetric wells, respectively. }
  \label{fgr:calc}
 \end{figure*}

Let us estimate the quantization energy in the triangular side wells (lh2 in the Fig.~\ref{fgr:calc}). For these states another approximation should be used: $\tilde{a}_B\gg L$ . For the raphe estimation we consider that the hole moves in a homogeneous electric field along z-axis. First let us find the value of the radius of this state $a$. The quantization energy counted from the bottom of the triangular well and the radius can be found from solving the equation: 

\begin{equation}
\label{f12}
{e^2 \over \varepsilon a }\approx 2.3{ \left[ \,\left({e^2 \over \varepsilon {a}^2}\right){\hbar \over \sqrt{2m_h}}\right]\,}^{2/3}.
\end{equation}

Solving this equation we get: $a\approx50~A,\hbar\omega\approx12~meV$. 

Figure~\ref{fgr:calc} shows the result of calculating the potential for holes in a symmetric and asymmetric structures. The calculation is performed by calculating the potential equation~\eqref{f8} by the finite element method.  

The energy of the electron at the lower quantization level in the symmetric quantum well, counted from the bottom of the CdTe well is $E_e=20~meV$. The energy of the electron in the asymmetric quantum well is slightly higher and is equal to  $E_e=23~meV$. 

{\begin{table} [h]
\small
  \caption{\ Energy of electrons and holes counted from the corresponding barriers. e1=25 meV; e2=0.1 meV.}
  \label{tbl4}
 \begin{tabular*}{\linewidth}{@{\extracolsep{\fill}}cccc}
    \hline
    hh1  & lh1 &  lh2 & hh2 \\
    \hline
    15 meV+e1 &	10 meV+e1 &	5 meV+e1 &	5 meV+e2 \\
      \hline
 \end{tabular*}
\end{table}}
 
\begin{widetext}

{\begin{table*} [h]
\small
  \caption{\ Energies of exciton transitions.  }
  \label{tbl5}
 \begin{tabular*}{\textwidth}{@{\extracolsep{\fill}}lcccc}
    \hline
     & e1-hh1& 	e2-hh2& 	e1-lh1& 	e1-lh2\\
    \hline
    Measured from the bottom of the valence band in CdZnTe &	40 meV&	7 meV&	35 meV&	30 meV \\
     \hline
    Interband transitions &	1.593 eV&	1.620 eV&	1.600 eV&	1.605 eV \\
      \hline
 \end{tabular*}
\end{table*}}

\end{widetext}

\section{Discussion}
\label{Disc}

The shape of the exciton reflection line allows us to determine the distance from the surface to the quantum well~\cite{9}. Indeed, the reflection coefficient $r$  from the structure containing the quantum well is: 

\begin{equation}
\label{f13}
r =  r_{01} + {t_{01} t_{10} {\rm e}^{2{\rm i}\varphi} \over 1- r_{10} r_{QW} {\rm e}^{2{\rm i}\varphi}}.
\end{equation}

Where transmittance coefficients at the vacuum-crystal interface on the crystal side is $t_{10}$  and on the vacuum side is  $t_{01}$;  $r_{01}$ and $r_{10}$  are reflection coefficients at the vacuum-crystal interface on the vacuum and crystal side; $r_{QW}$ is reflection coefficient from the quantum well, $\varphi=k \left (d+{L / 2}\right )$  is phase shift at the passage of light from the surface to the well, $k$  is wave vector of light. 

Neglecting the multiple reflections and the quadratic on $r_{QW}$ contribution, we obtain for the observed reflection coefficient:  

\begin{equation}
\label{f14}
R = {\vert r \vert}^2 \approx R_0 \left [ 1+ 2 {t_{01} t_{10}\over r_{01}} \left \lbrace r_{QW} {\rm e}^{2{\rm i}\varphi} \right \rbrace\right].
\end{equation}

Thus, the shape of the reflection spectrum is determined by the phase shift $\varphi$ . The thickness of the outer barrier in our structures is 90~nm, the refractive index in CdZnTe in this spectral region is $n= 2.65$. Hence, the run-up phase of the light wave when passing the barrier layer is equal to $\pi$  . That is, the shape of the exciton reflection contour from the well should be the inverse of the surface reflection contour, which can be seen in the spectra (Fig.~\ref{fgr:RefPL}~a). Thus, features of the reflection spectra at energies 1.597~eV, 1.602~eV, 1.609~eV, and 1.620~eV should be assigned to the exciton transitions in the quantum well.

The only difference between the asymmetric and symmetric structure is that in the CdMgTe barrier layer the refractive index at the energy of exciton resonances in the quantum well is  $n= 2.1$. In this case, the phase shift of the light wave as it passes through the cover layer is $\pi /2$ . This allows us to identify the spectral features at energies 1.604~eV, 1.610~eV, 1.620~eV, and 1.625~eV as belonging to the quantum well. These values are slightly higher than the energies of excitonic transitions in the symmetric quantum well. Indeed, for the asymmetric well the energy level should be slightly higher than for the symmetric well. In contrast to the well with symmetric barriers in the asymmetric structure, the shape of the exciton reflection line from the CdZnTe barrier is the same as that from the quantum well. Since this barrier is at the same distance from the surface, the phase shift of the light wave reflected from the well and from the CdZnTe layer when passing from the well to the surface is practically the same. 

Comparing the results obtained by numerical calculation, reflectance spectra, PL spectra and spectral dependences of Stokes parameters we obtain data   represented in Tables~\ref{tbl6} and \ref{tbl7}.

{\begin{table*} 
\small
  \caption{\ Symmetric well. Refractive index of light in the CdZnTe barrier $n=2.65$. The reflected light phase is reversed by 180 in comparison with reflecting from the surface. }
  \label{tbl6}
 \begin{tabular*}{\textwidth}{@{\extracolsep{\fill}}ccccc}
    \hline
    Phase [grad.]&	Reflectivity [eV]&	PL [eV]&	Absorption [eV]	&Transition  \\
          \hline
      180 &	1.597 &	1.597 	&1.596 &	e1-hh1 \\
      \hline
180 &	1.6017 &	no&	no 	&e1-lh1 \\
      \hline
180 &	1.6091 	&1.610& 	1.610& 	e1-lh2 \\
      \hline
0 &	1.615 &	no	&no	& not identified \\
      \hline
180 &	1.620 &	1.620 &	no&	e2-hh2 \\
      \hline
0 &	1.626 &	1.626 &	1.626 &	barrier\\
      \hline
 \end{tabular*}
\end{table*}}
{\begin{table*} 
\small
  \caption{\ Asymmetric well. Refractive index of light in the CdMgTe barrier is $n=2.1$. 
The phase is reversed by 90 degrees on reflection from the QW. The CdZnTe barrier has the same phase. }
  \label{tbl7}
 \begin{tabular*}{\textwidth}{@{\extracolsep{\fill}}cccc}
    \hline
    Phase [grad.]&	Reflectivity [eV]&	PL [eV]& Transition  \\
          \hline
      90 & not identified &		1.5986 &	trion \\
          \hline
90 &	1.6043 	&1.6047 &	e1-hh1 \\
          \hline

90 &	1.6111 	&1.6109 &	e1-lh1  \\
          \hline
90 &	1.6136 &	1.6126 & 	not identified 	 \\
          \hline
90 &	1.6209 &	1.6212 & e1-lh2 \\
          \hline
90 	&1.6251 &	no 	&e2-hh2 \\
          \hline
90 &	1.6313 &	1.6303& 	barrier \\
      \hline
 \end{tabular*}
\end{table*}}

First of all, the question arises: why some exciton states emerge as bright lines in the reflection spectra, but are completely absent in the PL spectra. Usually, it is just the opposite. Due to the energy relaxation of carriers, they accumulate in the lower energy states, from which the PL occurs. As a result, the intensity of the PL can be significant even at low oscillator strength of the optical transition. 

During non-resonant photoexcitation, holes rapidly lose energy by relaxing to the bottom of the corresponding bands. In our structure, the heavy holes~(hh) experience a quantum well of 2~meV depth whereas the light holes~(lh) in the CdTe layers have a 12.5~meV energy barrier height (Figure~\ref{fgr:calc}). From these layers, the light holes quickly move to the CdZnTe layers where their energy is lower. 

Electrons arrive at the bottom of their band much later than holes. This is confirmed by their greater mobility compared to holes and, therefore, a lower rate of energy loss. Once at the lower level of dimensional quantization in the QW, they begin to bind with holes to form excitons. First, highly excited bound states with large radius are formed. Electrons and holes, emitting acoustic phonons, and then optical ones, "descend" by energy to the ground exciton state~\cite{20,20i}.

Those light holes that are in the CdZnTe layers at the time of electron arrival to the QW form excitons (e1-lh2). But there are no holes in CdTe layers at the moment of electrons arrival and such excitons (e1-lh1) are not formed. 
Thus, the holes that got into the CdTe layers before the electrons arrived cannot form excitons and such excitons are not observed in the PL spectra. However, these excitons (e1-lh1) manifest brightly in the reflection spectra since they do not require an intermediate binding and energy relaxation process for their formation (Figure~\ref{fgr:RefPL}). 

	In asymmetric structures, this transition is visible in both the reflection and photoluminescence spectra. This is due to the fact that the high CdMgTe barrier effectively repels holes and does not let them escape. 
	
Another feature of the spectra studied is related to the phenomenon of birefringence caused by interface anisotropy. The spectra (Figure~\ref{fgr:conver}) show that birefringence occurs mainly on excitons with light holes. Indeed, it is the light holes that are in direct contact with the interfaces, while the heavy ones are mainly in the Coulomb field of the exciton (Figure~\ref{fgr:calc}). Their wave functions are of course somewhat distorted due to the anisotropy of the barriers, but they do not touch the interface directly. Therefore, the birefringence effect on them should be smaller. 

Let light, linearly polarized along axes  $x \| [100]$ and  $y \| [010]$, falls on the surface of the sample along axis $ z \| [001]$. In this case, degree of circular polarization of reflected light is 

\begin{equation}
\label{f15}
2iP_{circ} = r_{x'} r^\ast_{y'}-r_{y'} r^\ast_{x'},
\end{equation}
where axis  $x' \| [110] $ and  $y' \|[1\bm{\bar}{1}0]$ , $r_{x'}$  ,  $r_{y'}$ are amplitude reflection coefficients in the axes   $x'$ and $y'$. 

Degree of linear polarization of reflected light in the axes  $x'$ and $y'$ rotated by 45 degrees relative to the axes  $x$ and $y$: 

   		\begin{equation}
\label{f16}
2P_{lin'} = r_{x'} r^\ast_{y'}+r_{y'} r^\ast_{x'}.
\end{equation}

Degree of linear polarization of reflected light in the axes $x$ and $y$: 

   	\begin{equation}
\label{f17}
2P_{lin} = r_{x} r^\ast_{x}-r_{y} r^\ast_{y}.
\end{equation} 

Amplitude coefficient of exciton reflection from quantum well~\cite{22} 

   	\begin{equation}
\label{f18}
r_= {{\rm i}\Gamma_0 \over {\omega}_{0}-\omega - {\rm i}(\Gamma_{0}+\Gamma)},
\end{equation} 
here  $\Gamma_0 $ is radiative damping,  $\Gamma$ is non-radiative damping, $\omega_{0}$  is the exciton resonance frequency. 

Because of the interface anisotropy in heterostructures~\cite{7, 21}, the states of light and heavy holes at the interface are mixed. As a result, the reflection coefficients of light polarized along the axes $[110]$  and $[1\bm{\bar}{1}0]$  in the region of exciton resonances may differ strongly. The theory predicts that the radiative damping of the exciton $\Gamma_0 $  differs for these two directions, but all other exciton parameters  $\omega_{0}$ and  $\Gamma$ are the same~\cite{22}. Then, for the degree of circular polarization of the light reflected from the structure, we obtain: 

	\begin{equation}
\label{f19}
P_{circ}={\Gamma_{0}^{x'}\Gamma_{0}^{y'}\Delta\omega (\Gamma_0^{x'}-\Gamma_0^{y'}) \over {\vert\Delta\omega- {\rm i}\Gamma^{x'}\vert}^2 {\vert \Delta\omega- {\rm i}\Gamma^{y'} \vert}^2}.
\end{equation} 
Here $\Delta\omega=\omega_{0}-\omega$.

This formula shows that the degree of circular polarization $P_{circ} (\omega)$  is zero at the resonant frequencies of the exciton. For the degree of linear polarization in the axes $x'$ and $y'$:

	\begin{equation}
\label{f20}
P_{lin'}={\Gamma_{0}^{x'}\Gamma_{0}^{y'} \left [ (\Delta\omega)^2 + \Gamma^{x'}\Gamma^{y'} \right ] \over {\vert\Delta\omega- {\rm i}\Gamma^{x'}\vert}^2 {\vert \Delta\omega- {\rm i}\Gamma^{y'} \vert}^2} .
\end{equation} 

It can be seen that the degree of linear polarization in the rotated axes does not change the sign and tends to zero far from the exciton resonances. 

Let us estimate these quantities. From the experimental dependence (Figure~\ref{fgr:conver}) for amplitude of the linear polarization $P_{lin'} (\omega)$  we obtain: 
\begin{equation}
P_{lin'}= \left ({\Gamma_{0}^{x'} \over \Gamma^{x'}} {\Gamma_{0}^{y'} \over \Gamma^{y'}}\right )  \approx 0.15 .
\nonumber
\end{equation}
Hence $	{\langle \Gamma_{0} \rangle} / {\langle \Gamma \rangle} \approx 0.387 $ . That is, the non-radiation damping is about 3 times greater than the radiation damping. A similar value is obtained from the reflection spectrum. Indeed, the amplitude of the line in the reflection spectrum is $\sim	{\langle \Gamma_{0} \rangle} / {\langle \Gamma \rangle}  $ . 

The maxima and minima in the spectral dependence $P_{circ} (\omega)$ , are separated in frequency by the value $\Delta\omega~\approx~\Gamma $. Hence, we obtain: 
\begin{equation}
P_{circ} \approx {1 \over 2} \left [{(\Gamma_0^{x'})^2 - (\Gamma_0^{y'})^2\over \Gamma^{x'}\Gamma^{y'}} \right ]  \approx 0.15 .
\nonumber
\end{equation}
Hence $\Gamma_{0}^{x'} / \Gamma_{0}^{y'}\approx 2.5$. This value is close to the magnitude to that observed in ZnSe/BeTe type-II quantum wells~\cite{23}.

Such an anisotropy of the exciton radiation damping is caused by the mixing of light and heavy hole states caused by the low symmetry of a single interface $(\rm C_{2v})$ compared to the symmetry of the entire quantum well $(\rm D_{2d})$. As has been shown in~\cite{7, 20,20i}, interface anisotropy arises due to the orientation of chemical bonds in zinc-blend based structures. 

In type-II structures, the electron and hole are in different layers. In this case the exciton is localized directly at the interface itself, and the interface anisotropy manifests itself in the local properties of the structure. In our case, the anisotropy of the interfaces manifested in the anisotropy of the dielectric response, that is, in the macroscopic characteristic of the structure. 

In symmetric structures, the exciton wave function "touches" both barriers. Despite the low symmetry of each of them in the quantum well, in this case, the exciton "feels" the symmetry of the whole well, i.e., the $(\rm D_{2d})$ symmetry. In the asymmetric structure, the exciton feels stronger only one of the barriers having $(\rm C_{2v})$ symmetry, i.e., a dedicated axis [110] in the pit plane.

\section{Conclusion}
\label{Concl}
In the present study we investigated polarized reflection spectra from structures with quantum wells having symmetrical barriers Cd$_{0.9}$Zn$_{0.1}$Te/CdTe/Cd$_{0.9}$Zn$_{0.1}$Te and asymmetric Cd$_{0.9}$Zn$_{0.1}$Te/CdTe/Cd$_{0.4}$Mg$_{0.6}$Te. Stokes parameters of reflected light from these structures were measured. In the structures with symmetrical barriers, exciton resonances were detected in the reflection spectra, which do not show up in photoluminescence spectra. This peculiarity can be explained by the difficulty of population of some states at nonresonant optical excitation. 

In the structure with asymmetric barriers in the exciton resonance region, the phenomenon of light birefringence caused by reduced symmetry of the interfaces was found. This effect was observed mainly on the light exciton states, which is due to the fact that the wave functions of light holes is in direct contact with the interfaces, while the quantization of heavy holes is determined entirely by the Coulomb field of the electron.

The birefringence phenomenon is related to the atomic structure of interfaces in heterostructures based on semiconductors with zinc-blend structure, in which the covalent chemical bonds between atoms are aligned in the directions [110] and $[1\bm{\bar}{1}0]$. This results in a dedicated direction at the individual interface. The fact that an interface of infinitesimal thickness can determine the optical properties of the whole macroscopic structure is very surprising.

\acknowledgments 
This work was carried out with partial financial support from the Russian Science Foundation (project No. 21-12-00304). The authors would like to thank L.E.~Golub for helpful discussions.

\end{document}